\documentclass[10pt,a4paper,twoside]{article}
\usepackage{epsfig}
\usepackage{baltlat6}
\usepackage{array}
\newcommand{\rmd}{\mathrm{d}}               
\newcommand{\sauron}{\texttt{SAURON}}

\begin{document}
\ \
\vspace{0.5mm}
\setcounter{page}{000}
\vspace{8mm}

\titlehead{Baltic Astronomy, vol.\,00, 000--000, 2011}

\titleb{GRIGORI KUZMIN AND STELLAR DYNAMICS}

\begin{authorl}
\authorb{P.\ Tim de Zeeuw}{1,2} and
\authorb{Glenn van de Ven}{3}
\end{authorl}

\begin{addressl}
\addressb{1}{European Southern Observatory, Karl-Schwarzschild-Str.\ 2, 
             85748 Garching, Germany;  tdezeeuw@eso.org}
\addressb{2}{Sterrewacht Leiden, Leiden University, Postbus 9513, 
             2300 RA Leiden, The Netherlands}
\addressb{3}{Max-Planck Institute for Astronomy, K\"{o}nigstuhl 17, 
             69117 Heidelberg, Germany; glenn@mpia.de}
\end{addressl}

\submitb{Received: 2011 month dd; accepted: 2011 month dd}


\begin{summary} 
Grigori Kuzmin was a very gifted dynamicist and one of the towering
figures in the distinguished history of the Tartu Observatory. He
obtained a number of important results in relative isolation which
were later rediscovered in the West. This work laid the foundation for
further advances in the theory of stellar systems in dynamical
equilibrium, thereby substantially increasing our understanding of
galaxy dynamics.
\end{summary}

\begin{keywords}
stellar dynamics -- galaxies: elliptical and lenticular, cD --
galaxies: kinematics and dynamics -- galaxies: structure
\end{keywords}

\resthead{Grigori Kuzmin and Stellar Dynamics}
{de Zeeuw \& van de Ven}

\sectionb{1}{INTRODUCTION}

At the scientific symposium celebrating the 200th anniversary of the
Tartu Observatory, it is appropriate to consider the work of one of
its leading astronomers, Grigori G.\ Kuzmin. Kuzmin was born in 1917,
was a student at Tartu University from 1936 to 1940, and became an
assistant at the Observatory in the same year. He received his PhD
degree in 1952 and continued to work at the Observatory until after
his retirement in 1982. He passed away in 1988, having made
fundamental advances in the theory of galaxy dynamics.

In this contribution we concentrate on Kuzmin's work on galaxy models
with separable potentials, from spherical to axisymmetric to triaxial
(sections 2--4). The stellar orbits in these galaxy models all have
three exact integrals of motion which are quadratic functions of the
velocity components, allowing the calculation of many properties by
analytic means. The separable models not only provide insight into the
internal orbital structure and formation of galaxies, but they are
also essential for testing more sophisticated numerical modeling
approaches (section~5). Our conclusions follow in section 6.

\vfill\eject

\sectionb{2}{DYNAMICAL MODELS FOR GALAXIES}

In order to understand the dynamics of a collisionless stellar system
such as the Milky Way, the goal is to find the phase-space
distribution function $f(\vec{x}, \vec{v}) \geq 0$ which provides the
distribution of stars over all positions $\vec{x}$ and all velocities
$\vec{v}$. The mass-density distribution $\rho(\vec{x})$ follows by
integrating $f(\vec{x}, \vec{v})$ over the velocities: $\rho = \int f
\rmd^3 \vec{v}$. The velocities derive from the gravitational
potential $\Phi(\vec{x})$ via Newton's equation of motion ${\rm d}
\vec{v}/{\rm d}t = -\nabla \Phi(\vec{x})$. If the stellar system is
self-consistent, then the density is related to the potential by
Poisson's equation $4 \pi G \rho(\vec{x}) = \nabla^2 \Phi(\vec{x})$.

Traditionally two main approaches are taken to construct dynamical
models. Either one chooses a form for $f$ in order to find $\rho$,
which requires making a sensible Ansatz for $f$ based on some
pre-conceived notion on the galaxy formation process, or one assumes
$\rho$ and attempts to find $f$ by solving an integral equation. In
both cases it is advantageous to use Jeans' (1915) theorem which
states that $f$ depends on the six phase-space variables $\vec{x}$ and
$\vec{v}$ only through the single-valued (isolating) integrals of
motion admitted by the gravitational potential of the system. Since in
non-spherical systems there are generally at most three such
integrals, application of Jeans' theorem reduces the number of
variables in the problem of constructing a dynamical model.

The second order velocity moments $\langle v_i v_j\rangle$ of $f$
(with $i, j = 1, 2, 3$) are related directly to the density and the
potential by the equations of stellar hydrodynamics, also called the
Jeans equations. They can often be solved without knowledge of $f$,
but this generally requires additional assumptions about the shape of
the velocity distribution. Moreover, such Jeans models are not
guaranteed to correspond to a distribution function $f$ that is
everywhere non-negative, i.e., physical.

It is useful to consider spherical models. In this case the
Hamilton-Jacobi equation separates in standard spherical coordinates
$(r, \theta, \phi)$, and the motion of each star is governed by four
isolating integrals of motion: the energy $E$, and the three
components $L_x$, $L_y$ and $L_z$ of the angular momentum vector. All
orbits are planar rosettes (with the maximum angular momentum circular
orbit and the zero angular momentum radial orbit as extremes). The
mass distribution is defined by specifying the density profile
$\rho(r)$, and the associated gravitational potential $\Phi(r)$ can be
found by carrying out two single weighted integrations of
$\rho(r)$. There is only a single Jeans equation. Many self-consistent
models have been constructed by analytic means. These include
isotropic models with $f=f(E)$ found via an Abel inversion given the
density $\rho(r)$ (Eddington 1916), circular orbit models in which
only orbits with zero radial action are populated, the Osipkov
(1979)/Merritt (1985) models in which $f=F(E+aL^2)$ and many more
general distribution functions $f=f(E, L)$ with $L$ the total angular
momentum.

Over the past century many papers were written on the construction of
spherical models. Popular mass models are H\'enon's (1959) isochrone
and the $\gamma$-models (e.g., Dehnen 1993\footnote{Also derived by
  M.\ Franx in unpublished notes dated 1988.}), which include the
Jaffe (1982) and Hernquist (1990) models. Many of these were, in fact,
studied much earlier by Kuzmin and collaborators, in particular
Veltmann and later Tenjes. They derived density profiles, Jeans
solutions and distribution functions. The results were not well known
in the Western literature, but were summarized in English in Kuzmin
(1993). However, Kuzmin's main contribution lies elsewhere.

\vfill\eject

\sectionb{3}{KUZMIN'S PIONEERING WORK}

When Kuzmin started working on galaxy dynamics in Tartu in the early
nineteen forties, under very difficult circumstances, the field
concentrated on the Milky Way, for which many basic properties were
still poorly known.  This included the mass distribution, the
associated gravitational potential, the nature of the
three-dimensional stellar orbits other than those provided by the
epicyclic approximation for nearly-circular orbits, and the
phase-space distribution function. He contributed to all these topics.

It was known from classical mechanics that in a flattened axisymmetric
system such as the Milky Way, the energy $E$ and the angular momentum
component $L_z$ parallel to the symmetry axis [in cylindrical
  coordinates $(R, \phi, z)$], are integrals of motion. If the phase
space distribution function $f$ would depend only on $E$ and $L_z$ (by
the Jeans Theorem), then the second moments of the stellar motions
observed, e.g., near the Sun, should obey the relations $\langle v_R^2
\rangle = \langle v_z^2 \rangle$ and $\langle v_R v_z \rangle =
0$. But it was already clear from the observational data that $\langle
v_R^2 \rangle \ne \langle v_z^2 \rangle$ and $\langle v_R v_z \rangle
\ne 0$. This strongly suggested that the distribution function $f$ of
the Milky Way depends on a third parameter, or in other words that the
Galactic potential must support a third integral of motion $I_3$. The
nature of this integral was not well understood, and at the time only
an approximate expression been derived for nearly circular orbits with
small vertical harmonic oscillations (Lindblad 1933).

Non-spherical gravitational potentials with three exact integrals of
motion $E$, $I_2$ and $I_3$ had been known for half a century
(St\"ackel 1890, Eddington 1915, Clark 1937), but had received
relatively little attention. The integrals $I_2$ and $I_3$ are
quadratic functions of the velocity components. Chandrasekhar (1942)
assumed that $f=f(E+a I_2+b I_3)$ for such systems (the Ellipsoidal
Hypothesis), and attempted to derive the associated density $\rho$. He
demonstrated that the only self-consistent models satisfying this
assumed form for $f$ are spherical, and hence inconsistent with
flattened systems such as the Milky Way. Interest in such spherical
models was revived by Osipkov (1979) and Merritt (1985), but
Chandrasekhar's result did not encourage further work on the flattened
separable models. There was also little interest in pursuing the
opposite route: to try to solve for $f$ for densities $\rho$ that have
a separable potential. Van Albada (1952) (wrongly) demonstrated that
oblate axisymmetric separable potentials are not associated with mass
distributions that are everywhere non-negative, suggesting that their
usefulness for stellar dynamics was severely limited.

In the mean time, Kuzmin had in fact made very significant progress in
this area. He had worked on mass models for our Galaxy built by adding
non-homogenoeus ellipsoids, work later repeated by e.g., Burbidge,
Burbidge \& Prendergast (1962), had demonstrated that only isolating
integrals of motion should be used in Jeans' Theorem, well before
Lynden--Bell's (1962b) paper on the subject, and, most significantly,
had derived many interesting results for axisymmetric models with an
exact third integral of motion. The results were collected in his
seminal PhD Thesis (Kuzmin 1952\footnote{Translated into English by
  Tenjes and Einasto, including additions from 1969.}), and many of
these were published in the decade that followed.

Kuzmin considered axisymmetric mass models with a gravitational
potential
\begin{equation}
\label{eq:axipotsep}
\Phi_S = \frac{F(\lambda)-F(\nu)}{\lambda-\nu},
\end{equation}
where $(\lambda,\phi,\nu)$ are prolate spheroidal coordinates, with
$(\lambda, \nu)$ the two roots of
\begin{equation}
\frac{R^2}{\tau+\alpha} + \frac{z^2}{\tau+\gamma} = 1,
\end{equation}
and the constants $\alpha$ and $\gamma$ such that $-\gamma \leq \nu
\leq -\alpha \leq \lambda$ and $F(\tau)$ an arbitrary smooth function
$(\tau=\lambda, \nu)$. Surfaces of constant $\lambda$ are spheroids,
and surfaces of constant $\nu$ are hyperboloids of two sheets,
respectively. In these potentials, the Hamilton-Jacobi equation
separates in $(\lambda, \phi, \nu)$, resulting in three exact integrals
of motion $E$, $L_z$ and $I_3$, which are quadratic in the velocities
and can be written explicitly. Kuzmin showed that such potentials have
useful associated densities, given by a simple formula, and that the
density $\rho(R,z) \ge 0$ if and only if $\rho(0, z) \ge 0$, a result
now known as Kuzmin's theorem (de Zeeuw 1985b).

Kuzmin assumed a simple form for the density profile along the short
axis, $\rho(0, z)$, where $z^2=\tau+\gamma$, which in the notation
used here would be equivalent to:
\begin{equation}
\label{eq:powerlawtau}
\frac{\rmd}{\rmd \tau} (\tau+\gamma)^{1/2} F^\prime(\tau) \propto \tau^{-n/2}.
\end{equation}
He showed that the case $n=3$ provides a fair approximation to the
Milky Way potential as it was known at the time (without dark halo);
the model is nearly spheroidal and is in fact a flattened
generalization of H\'enon's isochrone. The density (and potential) can
be written explicitly in terms of $R$ and $z$, avoiding the need to
use spheroidal coordinates.  The case $n=4$ corresponds to an exactly
spheroidal model $\rho(R, z)=\rho_0/(1+m^2)^2$ with $m^2=R^2/a^2+z^2/c^2$
and $\alpha =-a^2, \gamma=-c^2$. In the limit of extreme flattening
both the $n=3$ and the $n=4$ models reduce to an infinitesimally thin
disk with surface density $\Sigma(R)=\Sigma_0/(1+R^2/a^2)^{3/2}$. This is
the Kuzmin (1953) disk which was rediscovered by Toomre (1963). It has
many remarkable properties, including the fact that the potential
above the disk is equal to that of a point mass at $(R=0, z=-a)$ below
the disk and the potential below the disk is that of a point mass at
$(R=0, z=+a)$ above the disk.  Kuzmin furthermore noted that the model
with $n=n_0$ is the weighted sum of models with $n>n_0$; this built on
his pioneering 1943 work on construction of models by superposition of
inhomogeneous spheroids.

The stellar orbits in oblate separable models are bounded by the
orthogonal confocal coordinate surfaces, and are now called short-axis
tubes. Their properties are similar to those of the orbits in the
Milky Way found numerically by Ollongren (1962) using Schmidt's (1956)
mass model. Kuzmin \& Kutuzov (1962) calculated the $f(E, L_z)$ for
the model $n=3$. They realized that $\rho(R, z)$ can be written
explicitly as $\rho(R, \Phi)$ without any reference to spheroidal
coordinates, which allows computing $f(E, L_z)$ via series expansions
\'a la Fricke (1952).  This work occurred in parallel with
Lynden-Bell's (1962a) construction of self-consistent flattened
Plummer models using integral transforms. But Kuzmin \& Kutuzov
apparently did not investigate three-integral distribution functions
for oblate (and the related prolate) density distributions. This was
left to Dejonghe \& de Zeeuw (1988) who made full use of the many
elegant properties of the Kuzmin-Kutuzov mass model, derived a closed
form for $f(E, L_z)$ (see also Batsleer \& Dejonghe 1993) as well as a
number of three-integral distribution functions, and computed the
projected surface density and the observed kinematic properties.

\vfill\eject

\sectionb{4}{TRIAXIAL GALAXY MODELS}

Kuzmin also generalized much of his work on oblate separable models to
those with triaxial shapes. This work is largely unpublished except
for a very concise four-page summary (Kuzmin 1973) in the proceedings
of the Alma Ata conference in 1972, which was translated into English
in Kuzmin (1987).  The paper considers triaxial models with potentials
separable in confocal ellipsoidal coordinates $(\lambda, \mu, \nu)$
defined as the three roots for $\tau$ of
\begin{equation}
\frac{x^2}{\tau+\alpha} + \frac{y^2}{\tau+\beta} + \frac{z^2}{\tau+\gamma} = 1,
\end{equation}
with $(x, y, z)$ the usual Cartesian coordinates, and with constants
$\alpha, \beta, \gamma$ such that $-\gamma \leq \nu \leq -\beta \leq
\mu \leq -\alpha \leq \lambda$. For potentials of the form
\begin{equation}
\Phi_S = \frac{F(\lambda)}{(\lambda-\mu)(\lambda-\nu)} 
      +\frac{F(\mu)}{(\mu-\nu)(\mu-\lambda)} 
      +\frac{F(\nu)}{(\nu-\lambda)(\nu-\mu)},
\end{equation}
all orbits have three exact integrals of motion $E$, $I_2$ and $I_3$,
quadratic in the velocities. Kuzmin noted that the orbits can be
divided into four main families, that the density associated to
$\Phi_S$ is non-negative, i.e., $\rho(x,y,z) \ge 0$ if and only if
$\rho(0,0,z) \ge 0$, and that it can be expressed elegantly via a
`Kuzmin formula'.  The set of models includes an exactly ellipsoidal
model, also known as the `perfect ellipsoid':
\begin{equation}
\rho(x, y, z)=\rho_0/(1+m^2)^2, \qquad \hbox{\rm with} \qquad
              m^2=x^2/a^2+y^2/b^2+z^2/c^2,
\end{equation}
where $\gamma=-c^2$, $\beta=-b^2$ and $\alpha=-a^2$. The four major
orbit families are referred to as box orbits, inner long-axis tube
orbits, outer long-axis tube orbits, and short-axis tube orbits, and
are identical to the four major orbit families found in
Schwarzschild's (1979) numerical model for a stationary triaxial
galaxy with a constant density core (de Zeeuw 1985a).

All these results for triaxial models were obtained via an independent
route by the first author in the period 1982--1985, following a
suggestion by Lynden-Bell to `look into how well separable potentials
could be made to fit a general potential in the finite density core of
a galaxy' (de Zeeuw \& Lynden--Bell 1985). This lead to the discovery
of the perfect ellipsoid (eq.\ 6) in a fairly roundabout way --- as
the ellipsoidal model for which the series expansion of the potential
fits a separable one to all orders --- and to the subsequent
rediscovery of the Kuzmin (1956) paper on oblate separable models. The
latter inspired the independent derivation of the Kuzmin formula for
the triaxial case, and was followed by the full derivation of the
gravitational potential for oblate, prolate and triaxial models given
as the sum of two weighted single integrations of the density profile
along the short axis (de Zeeuw 1985b; de Zeeuw, Peletier \& Franx
1986). It was Jaan Einasto who cheerfully pointed out to the first
author in 1985 during the Princeton IAU Symposium on Dark Matter that
`Kuzmin had already done it all for triaxial models many years
earlier'. This overstated the case a little, as the derivation of the
potential given the short-axis density profile was genuinely new. But
the trail led to the Alma Ata conference contribution already
mentioned, and the decision to have it translated into English and
included as an Appendix of the proceedings of IAU Symposium 127
(Kuzmin 1987). The text was carefully checked by Kuzmin himself about
one year before he passed away.

As Kuzmin knew, each separable triaxial model can be written as the
weighted integral of constituent perfect ellipsoids, with the weight
function found by means of a Stieltjes transform. It follows that the
projected surface density is the same weighted integral of the
projections of the constituent ellipsoids, which are elliptic
disks. This property led to a new and powerful method to find the
three-dimensional gravitational potential of non-axisymmetric disks
(Evans \& de Zeeuw 1992). Franx (1988) had already deduced general
properties of the projections of the separable models. All these
properties are in fact shared by a much larger set of models: each
ellipsoid $\rho=\rho_0/(1+m^2)^p$ with $p=n$ or $p=n/2$ generates a
similar family (de Zeeuw \& Pfenniger 1988; Evans \& de Zeeuw 1992),
but the motion is separable only for the models generated by the $p=2$
ellipsoids.

The second moment tensor is diagonal for the separable models, and the
diagonal elements are related to $\rho$ and $\Phi_S$ by the Jeans
equations, which are three partial differential equations for three
unknowns (Lynden-Bell 1960). Kuzmin and others had derived special
solutions for the axisymmetric and elliptic disk limiting cases, but
the general equations for triaxial geometry resisted solution for more
than forty years, until van de Ven et al.\ (2003) found it by
classical methods. As for all Jeans solutions, they are not guaranteed
to correspond to a physical model with $f \ge 0$, but do allow
efficient calculation of the second velocity moments.

It has also been possible to construct analytic self-consistent
models, but the existence of more than one major orbit family, in fact
four, implies that $f(E, I_2, I_3)$ is not uniquely defined by
$\rho(\lambda, \mu, \nu)$, and certain assumptions have to be made.
Populating only the tube orbits with zero radial action (thin tubes)
together with the box orbits, Hunter \& de Zeeuw (1992) constructed
models that can be considered as the triaxial generalization of the
circular orbit spheres.  Van de Ven et al.\ (2008) gave a
comprehensive description of the so-called Abel models in which $f =
\sum_i f_i(E + a_i I_2 + b_i I_3)$, with $a_i$ and $b_i$ constants for
each component $i$. These had been studied earlier by Dejonghe \&
Laurent (1991) and Mathieu \& Dejonghe (1999), and are triaxial
generalizations of the spherical Osipkov--Merritt models.\looseness=-2

The above brief summary shows that through Kuzmin's work and the
subsequent follow-up, the theory of stationary triaxial dynamical
models is now as comprehensive as that for spheres. This also means
that by matching these models with observations of galaxies --- such
as their observed surface brightness to infer the intrinisic density,
and their observed stellar kinematics to constrain the gravitational
potential and internal velocity distribution --- we can gain insight
in the internal dynamics of a much larger range of non-spherical
stellar systems, including early-type galaxies.

\sectionb{5}{EARLY-TYPE GALAXIES}

Measurements of the stellar kinematics of early-type galaxies with
integral-field spectrographs such as \sauron\ (Bacon et al.\ 2001) show
that they can be divided into \emph{slow rotators} and \emph{fast
  rotators}, different from their morphological classification in
ellipticals and lenticulars (Emsellem et al.\ 2007, 2011).  Most fast
rotators, including lenticular as well as many elliptical galaxies,
seem to be close to oblate axisymmetric disk-like systems (Cappellari
et al.\ 2007). On the other hand, various slow-rotator elliptical
galaxies show clear deviations from axisymmetry in their photometry
such as isophotal twists, as well as in their kinematics such as
kinematic misalignment and kinematically decoupled components (see
e.g.\ Emsellem et al.\ 2004, and Fig.~1), indicating that they are
triaxial.

The separable triaxial galaxy models described above are able to
capture much of the rich internal dynamics of these slow-rotator
galaxies.  However, the presence of central density cups and
super-massive black holes imply that there are no \emph{global}
integrals $I_2$ and $I_3$. The tube orbits are relatively unaffected,
but the box orbits are replaced by a mix of so-called boxlets
(higher-order resonant orbits) and chaotic orbits (e.g., Gerhard \&
Binney 1985).  While separable models are very useful to provide
insights as well as tests (e.g., van de Ven et al.\ 2008), to model
these elliptical galaxies in detail, a more general technique such as
Schwarzschild's (1979) numerical orbit superposition method is needed.

Schwarzschild's method allows for an arbitrary gravitational
potential, with possible contributions from a central black hole and a
dark matter halo. The equations of motions are integrated numerically
for a representative library of orbits. Next, the orbital weights are
determined for which the combined density and higher order velocity
moments best fit simultaneously the observed surface brightness and
(two-dimensional) kinematics. The resulting distribution of (positive)
orbital weights represents the distribution function (Vandervoort
1984), which is, by construction, guaranteed to be non-negative
everywhere.  Our recent implementation in triaxial geometry (van de
Ven et al.\ 2008, van den Bosch et al.\ 2008, van den Bosch \& van de
Ven 2009), allows the construction of accurate models for slow-rotator
galaxies and to establish their internal phase-space structure.

\begin{figure}[t]
\includegraphics[width=1.0\textwidth]{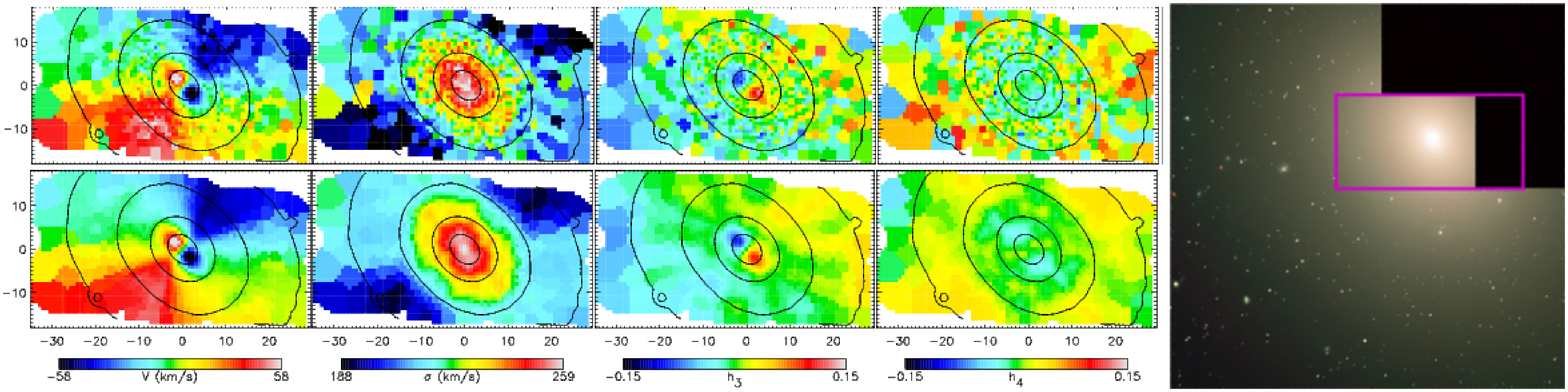}
\captionb{1}{Triaxial dynamical model of the elliptical galaxy
  NGC\,4365. \emph{Left:} Two-dimensional stellar kinematic
  observations of NGC\,4365 obtained with the integral-field
  spectrograph \sauron\ (top row) accurately fitted with our numerical
  implementation of Schwarzschild's orbit superposition method in
  triaxial geometry (bottom row). From left to right: mean
  line-of-sight velocity and dispersion, and higher-order
  Gauss-Hermite velocity moments, which quantify deviations from a
  Gaussian velocity profile. \emph{Right:} The black contours drawn on
  the maps are the isophotes derived from Hubble Space Telescope
  imaging of NGC\,4365 shown on the right-hand side with a magenta box
  indicating the extent of the \sauron\ observations. Whereas the
  imaging shows a smooth elliptical luminosity distribution, the
  kinematics reveal a complex non-axisymmetric orbital distribution
  with an (apparent) kinematically decoupled core.}
\end{figure}

One of these galaxies is NGC\,4365, a giant E3 galaxy with a core that
appears to rotate around the short axis (Surma \& Bender 1995; Davies
et al.\ 2001), almost perpendicular ($\sim$82\dgr) to the main body
rotation around the long-axis (Fig.~1). The customary interpretation
has been that this core is kinematically distinct (a so-called KDC)
and a remnant of a last major accretion $\sim$12\,Gyr ago.  However,
triaxial Schwarzschild models fitted to the \sauron\ stellar
kinematics (van den Bosch et al. 2008; Fig.~1), revealed that the KDC
is part of the single triaxial structure: short axis tube orbits
dominate throughout, but half of them counter-rotate leading to
vanishing net rotation around the short-axis, except in the center due
to a small in-balance leading to the appearance of a KDC. The
intrinsic shape changes from oblate triaxial in the central regions to
prolate triaxial in the outer parts where long-axis tube orbits
provide the net rotation seen in the observed velocity field. 

Triaxial Schwarzschild models for another dozen of these slow-rotator
ellipticals reveal a similar picture of galaxies dominated by short
axis tube orbits with canceling rotation apart from a small in-balance
in the center leading to the KDCs. In addition, these slow-rotator
galaxies contain on average 20\% box orbits consistent with their
mildly triaxial shape (van den Bosch et al., in prep.).  Several
fast-rotator galaxies also contain KDCs, but typically a magnitude
smaller in size, and they seem to be truly decoupled axisymmetric
disk-like components with a younger stellar population (see also
McDermid et al.\ 2006).  The slow-rotator galaxies with these {\sl
  apparent} KDC's, in harmony with their smooth age and metallicity
gradients, provide a true challenge for (cosmological) merger
simulations: Dry mergers over-predict the fraction of box orbits
(e.g., Jesseit et al.\ 2005) but invoking gas seems to require special
conditions to make the merger remnants as round as the observed
slow-rotator ellipticals (e.g., Hoffman et al.\ 2011, Bois et
al.\ 2011).

\sectionb{6}{CONCLUDING REMARKS}

Kuzmin also contributed to other areas of galactic dynamics, and
trained a number of students, amongst which Einasto, Kutuzov and
Veltman. The relative isolation in which he worked resulted in a
delayed recognition of many of his achievements in the West. This was
due in part to limited possibilities for publishing, information
exchange and travel in the Soviet era, and in part to provincialism,
as few in the West were able to read Russian. Kuzmin's seminal 1956
paper was published in the Russian Astronomical Journal in the last
year before it started to be translated in English. His papers often
had a short English summary attached, and he sent English synopses of
his work to key astronomers, amongst which Lynden--Bell, Oort and
others, but these were not widely distributed. Perek's (1962) review
did help advertize his results, but even so, some of his work was
independently rediscovered. Today's on-line availibility of preprints
and journals, including new and old Baltic Astronomy publications, and
the strongly improved possibilities for travel, should go a long way
towards avoiding that genuine discoveries are overlooked and repeated.

A number of the ideas in Kuzmin's (1956) paper on oblate separable
models were generalized to prolate and triaxial models and in turn led
to further developments, as described in sections 3 and 4.  These
contributed significantly to bringing the modeling of separable
triaxial systems to the level reached for spherical systems: a theory
of mass models and their projections, full understanding of the
orbits, general solution of the Jeans equations, and construction of
anisotropic distribution functions. This effort is valuable for
testing more sophisticated numerical triaxial models and for gaining
insight into the dynamical structure of early-type galaxies. As such,
Kuzmin's work contributed substantially to our understanding of galaxy
dynamics. He belongs in the list of very famous astronomers associated
with Tartu Observatory.


\thanks{The authors are grateful to Jaan Einasto and Peeter Tenjes for
  helpful discussions and for a preview of the English translation of
  Kuzmin's PhD thesis and subsequent addenda, and to Anu Reinart and
  Laurits Leedj\"arv for generous hospitality at the Tartu
  Observatory.}


\References

\refb Bacon R., et al., 2001, MNRAS, 326, 23

\refb Batsleer P., Dejonghe H., 1993, A\&A, 271, 104

\refb Bois M., et al., 2011, MNRAS, in press (arXiv:1105.4076)

\refb Burbidge E.M., Burbidge G.R, Prendergast K.H., 1959, ApJ, 130, 739
 
\refb Cappellari M., et al., 2007, MNRAS, 379, 418

\refb Chandrasekhar S., 1942, Principles of Stellar Dynamics, Univ. of
      Chicago Press

\refb Clark G.L., 1937, MNRAS, 97, 182

\refb Davies R.L., et al., 2001, ApJ, 548, L33

\refb Dehnen W., 1993, MNRAS, 265, 250

\refb Dejonghe H., de Zeeuw P.T., 1988, ApJ, 333, 90
 
\refb Dejonghe H., Laurent D., 1991, MNRAS, 252, 606
 
\refb de Zeeuw P.T. 1985a, MNRAS, 216, 273

\refb de Zeeuw P.T. 1985b, MNRAS, 216, 599 

\refb de Zeeuw P.T., Lynden--Bell D., 1985, MNRAS, 215, 713 

\refb de Zeeuw P.T., Peletier R.F., Franx M., 1986, MNRAS 221, 1001

\refb de Zeeuw P.T., Pfenniger D., 1988, MNRAS, 235, 949

\refb Eddington A.S., 1915, MNRAS, 76, 37

\refb Eddington A.S., 1916, MNRAS, 76, 572

\refb Emsellem E., et al., 2004, MNRAS, 352, 721
 
\refb Emsellem E., et al., 2007, MNRAS, 379, 401

\refb Emsellem E., et al., 2011, MNRAS, 414, 888

\refb Evans N.W., de Zeeuw P.T., 1992, MNRAS, 257, 152

\refb Franx M., 1988, MNRAS, 231, 285

\refb Fricke W., 1952, AN, 280, 193

\refb Gerhard O.E., Binney J.J., 1985, MNRAS, 216, 467

\refb H\'enon M., 1959, AnAp, 22, 126
  
\refb Hernquist L., 1990, ApJ, 356, 359

\refb Hoffman L., Cox T.J., Dutta S., Hernquist L., 2010, ApJ, 723, 818

\refb Hunter C., de Zeeuw P.T., 1992, ApJ, 389, 79

\refb Jaffe W., 1983, MNRAS, 202, 995

\refb Jeans J.H., 1915, MNRAS, 76, 70

\refb Jesseit R., Naab T., Burkert A, 2005, MNRAS, 360, 1185

\refb Kuzmin G.G., 1952, PhD Thesis, Tartu University

\refb Kuzmin G.G., 1953, Tartu Astro. Obs. Teated, 1

\refb Kuzmin G.G., 1956, Astr. Zh., 33, 27

\refb Kuzmin G.G., 1973, Proc.\ All-Union Conf.\ Dynamics of Galaxies
      and Clusters, Alma Ata, ed.\ T.B.\ Omarov, 71

\refb Kuzmin G.G., 1987, IAU Symposium 127, 553-556

\refb Kuzmin G.G., 1993, IAU Symposium 153, 363-366

\refb Kuzmin G.G., Kutuzov S.A., 1962, Bull.\ Abastumani Astroph.\ Obs, 27, 82

\refb Lindblad B., 1933, Handbuch der Astrophysik (Springer Verlag), 2, 1033

\refb Lynden--Bell D., 1960, PhD Thesis, Cambridge University

\refb Lynden--Bell D., 1962a, MNRAS, 123, 447 

\refb Lynden--Bell D., 1962b, MNRAS, 124, 1 


\refb Mathieu A., Dejonghe H., 1999, MNRAS, 303, 455

\refb McDermid R.M., et al., 2006, MNRAS, 373, 906

\refb Merritt D.R., 1985, AJ, 90, 1027

\refb Ollongren A., 1962, BAN, 16, 241

\refb Osipkov L., 1979, SOv.\ Astron.\ Letters, 5, 42

\refb Perek L., 1962, Advances in Astronomy and Astrophysics, 1, 165

\refb Schmidt M., 1956, BAN, 13, 15

\refb Schwarzschild M., 1979, ApJ, 232, 236

\refb St\"ackel P., 1890, Math. Ann., 35, 91 

\refb Surma P., Bender R., 1995, A\&A, 298, 405

\refb Toomre A., 1963, ApJ, 138, 385

\refb van Albada G.B., 1952, Proc.\ KNAW, B55, 5, 620

\refb van de Ven G., Hunter C., Verolme E.K., de Zeeuw P.T., 2003,
MNRAS, 342, 1056

\refb van de Ven G., de Zeeuw P.T., van den Bosch R.C.E., 2008, MNRAS, 385, 614
 
\refb van den Bosch R.C.E., van de Ven G., Verolme E.K., Cappellari
M., de Zeeuw P.T., 2008, MNRAS, 385, 647
 
\refb van den Bosch R.C.E., van de Ven G., 2009, MNRAS, 398, 1117
 
\refb Vandervoort P.O., 1984, ApJ, 287, 475


\end{document}